\documentclass[twocolumn,seceq]{jpsj2} 
\usepackage{latexsym}
\usepackage{color}

\newcommand{\nn}{\nonumber \\}
\newcommand{\udarrow}[2]{\smash{\mathop{%
  \hbox to 0.4cm{$\rightleftharpoons$}}\limits^{#1}\limits_{#2}}}
\newcommand{\uarrow}[1]{\smash{\mathop{%
  \hbox to 0.4cm{$\rightarrow$}}\limits^{#1}}}

\title{%
Chirality Selection in Open Flow Systems and in Polymerization
}

\author{%
Yukio \textsc{Saito}\thanks{yukio@rk.phys.keio.ac.jp}
 and Hiroyuki \textsc{Hyuga}\thanks{hyuga@rk.phys.keio.ac.jp}
}

\inst{%
Department of Physics, Keio University, Yokohama 223-8522 
}

\recdate{\today}

\abst{%
As an attempt to understand the homochirality of organic molecules in life,
a chemical reaction model is proposed where the production
of chiral monomers from achiral substrate is catalyzed 
by the polymers of the same enatiomeric type. 
This system has to be open because in a closed system the 
enhanced production of chiral monomers by enzymes is compensated by
 the associated enhancement in back reaction, 
 and the chiral symmetry is conserved.
Open flow without cross inhibition is shown to lead
to the chirality selection in a general model.
In polymerization, the influx of substrate from the ambience
and the efflux of chiral products for purposes other than
the catalyst production make the system necessarily open.
The chiral symmetry is found to be broken if the influx of substrate
lies within a finite interval.
As the efficiency of the enzyme increases, the maximum value of the
enantiomeric excess approaches unity so that the chirality selection
becomes complete.
}

\kword{%
homochirality, open system, polymerization,  enantiomer,
rate equation, dynamical chiral symmetry breaking
}

\begin{document}
\sloppy
\maketitle

\section{Introduction}
Between two possible stereostructures 
of organic molecules, 
i.e. 
a right-handed (D) and a mirror-image left-handed (L) form, 
life on earth has chosen only one type:
L-amino acids and D-sugars.
\cite{feringa+99}
Various mechanisms for the origin of this chiral asymmetry, 
so called homochirality, 
have been proposed,
\cite{feringa+99}
but the predicted asymmetry turned out to be very minute, 
and therefore, it has to be amplified to explain the homochirality. 

Frank showed theoretically that an autocatalytic production 
of chiral molecules with an antagonistic process
amplifies enantiomeric excess (ee)
and brings about the homochirality.\cite{frank53}
His theory has had no experimental support for a long time.
Recently, an example of ee amplification has been found
in the production of pyrimidyl alkanol,
\cite{soai+95}
and the temporal evolution was explained by the second-order autocatalytic
reaction.\cite{sato+03}
In our previous papers\cite{saito+04a,saito+04b,saito+04c,saito+05a}, 
we have reported that, in addition to the nonlinear autotacalysis,
recycling of achiral substrates by the decomposition of chiral products 
accomplishes complete homochirality in a closed system
without any antagonistic processes.

The chemical reaction model we proposed
\cite{saito+04a}
involves production 
of a chiral molecule C from achiral molecules A and B.
With an assumption that there is an ample amount of the material B, 
the reaction is governed solely by the concentration $a$ of the substrate A. 
We adopt the convention of $R,S$ representation to denote 
two enantiomers of the chiral product C as ($R$)-C and ($S$)-C,
which are further 
abbreviated 
as R and S, respectively, for brevity. 
The reaction is assumed to include a
 nonlinear autocatalytic effect as well as a recycling back reaction,
and the concentrations $r$ and $s$ of chiral species 
R and S develop by the following rate equations;
\begin{align}
 \dot r = (k_0+ k_1 r + k_2 r^2) a - \lambda_0 r,
\nonumber \\
 \dot s = (k_0+ k_1 s + k_2 s^2) a - \lambda_0 s,
\label{eq1}
\end{align}
where $\dot x$ represents the time derivative $\dot x= dx/dt$.
The reaction system is assumed to be closed so that
 the concentration $a$ of the substrate A is determined by the
conservation as $a=c_0 - r-s$. Here, the constant $c_0$ is
fixed at the initial time as $c_0=a_0+r_0+s_0$.
Reaction coefficients $k_0,~k_1$ and $k_2$ correspond to those 
in the production
processes 
without autocatalysis, with a linear and  a
quadratic autocatalysis, respectively.
$\lambda_0$ is a rate coefficient of decomposition 
process  from the product R or S to the substrate A.
The major conclusion we have drawn out from the model eq.(\ref{eq1})
is that, only in the case with a quadratic autocatalysis and decomposition 
process ($k_2,~\lambda >0$), 
the chiral symmetry breaking is possible.
\cite{saito+04a}

However, the relevance of pyrimidyl alkanol to life is suggestive at best, 
and the problem of the
homochirality in life is still not yet resolved.
One way to address this problem is to construct a model
that incoporates some characteristic features of organic molecules in life.
For instance, amino acids
have the potential to polymerize into chain molecules,
as peptides and proteins. They act 
as catalysts or enzymes to produce various molecules, 
and among them may be amino acids themselves, although 
this expectation has not so far been confirmed experimentally.  
Based upon this hypothesis,
Sandars\cite{sandars03} proposed a "toy model" for the generation 
of homochirality such 
 that polymers catalyze production process of chiral monomers
from achiral substrates. 
He included cross inhibition effect such that the 
addition of wrong handed enantiomer to the polymer halts 
further polymerization\cite{sandars03,brandenburg+04}.
The cross inhibition is similar to the mutual destruction effect 
introduced by Frank.
In this paper, we would like to 
propose an alternative model for the generation of homochirality,
exploiting a chracteristic feature that reactions take place
in an open system; we do not require processes of cross inhibition.

Most of amino acids (19 out of 20 species ) are chiral,  
but they are composed of achiral molecules as carbon dioxides, 
anmonium, water and so on.
These achiral substrates are supplied from and leave away to the ambience.
On the other hand, the products, amino acids, 
polymerize to form peptides and proteins, which not only function as 
catalysts or enzymes but also are  used 
for many other purposes as body construction, metabolism, etc.
This means that some amino acids leak away from the reaction path of
catalysts production:
The chiral products thus flow out from the reaction system in concern.
All these features lead us to the necessity to study an open system. 
We construct a minimal reaction model including the above ingredients.
In a certain limit, we can show that open reaction systems are reducible 
to our previous model eq.(\ref{eq1}) or similar models 
in a closed system.\cite{saito+04a}

In the following section \S 2, a simple model of chiral molecule
production under an open flow 
is shown to behave similar to that in a closed system after
a transient time.
In \S 3, the polymerization model is studied, and the open flow is
found necessary to break the chiral symmetry for this model.
The results are summarized in the last section \S4. 
Some details of stability analysis is relegated in the appendix.

\section{Chemical Reaction in an Open System under Flow}

First we extend our nonlinear autocatalytic system proposed previously
\cite{saito+04a}
to an open system under flow, 
and show that a flow plays essentially the same role as 
the recycling process in a closed system.

As for the origins of the incoming flow,
substrate molecules A may be produced
by some chemical reactions inside the system, or there may be a supply
of A's by a flow from ambience. In any case, 
the incoming flux of A is assumed to be constant and is denoted as $F$.
On the other hand, dissipation may be due to the outgoing
flow which takes away part of all the chemical species
from the region where the reactions are taking place.
For simplicity the rate of outflow is assumed common to all chemical 
species, and denoted as $\mu_0$.
Combining the effects of autocatalytic chemical reaction and in- and 
out-flows,
the rate equations are written as
\begin{align}
\dot r &= f(r) a - \lambda_0 r- \mu_0 r,
\nonumber \\
\dot s &= f(s) a - \lambda_0 s - \mu_0 s,
\nonumber \\
\dot a &= -(f(r)+f(s)) a + \lambda_0 (r+ s) +F - \mu_0 a.
\label{eq2}
\end{align}
Here $f(r)=k_0+k_1 r + k_2 r^2$ is the production rate of R
including the effect of autocatalysis, $f(s)$ is the corresponding one for S,
$\lambda_0$ is the rate of back reaction from chiral products R or S to
the substrate A.
It is then easy to show that the sum of all the species relax 
to a constant value
$c_0=F/\mu_0$ in a relaxation time $\mu_0^{-1}$ as
$r+s+a=c_0+(r_0+s_0+a_0-c_0)\exp(-\mu_0t)$.
Thus, after the relaxation time $\mu_0^{-1}$, the open system we consider
reduces effectively to a closed system (\ref{eq1}) at a fixed total 
concentration $c_0$ with a modified back reaction rate 
$\lambda= \lambda_0+\mu_0$.
The difference between the two systems is essentially the value of $c_0$;
in a closed system it is determined
as an initial condition, whereas in an open system it is controlled
by the in- and out-going flows.
Therefore,
 even in the case that the back reaction is absent ($\lambda_0=0$),
 the outflow plays essentially the same role as the back reaction, 
 and the chirality selection can be realized in combination
 with nonlinear autocatalysis.
This conclusion leads us to propose an experiment of the Soai reaction
\cite{soai+95} 
under the flow with an expectation 
that the complete homochirality can be realized.

The analysis in this section reveals 
that the
generic form of the model (\ref{eq1}) 
has a wide range of applicability to
describe the chirality selection in various systems. 
In the following section
in the analysis of polymerization processes, 
we encounter a generalized form of this type of rate equations
\begin{align}
\dot r&=f(r)a-\lambda(r)r\nonumber\\
\dot s&=f(s)a-\lambda(s)s,
\label{eq3}
\end{align}
where 
$f(r)$ is an effective  production rate and 
$\lambda(r)\,$ is an effective rate of back reaction.
The substrate concentration $a$ follows 
another generic equation.
Equation (\ref{eq3}) does not contain the cross inhibition term 
such as $-\lambda_c rs\,$.
By suitable conditions for the functions  $ f(r),~\lambda(r)$
and other parameters, 
the chirality selection is realized as is demonstrated in the following. 
Some of the details  is described in the Appendix.

\section{Chirality Selection in Polymerization}

We now study the chiral symmetry breaking in 
the catalyst
polymerization processes.
Our model starts from chemical reactions 
of chiral molecule production from achiral substrate, as
\begin{align}
A ~ \udarrow{k_0}{\lambda_0} R, \qquad
A ~ \udarrow{k_0}{\lambda_0} S . 
\label{eq4}
\end{align}
Here an achiral substrate A reacts to produce 
chiral species R and S with a rate $k_0$ and the reverse reaction
takes place with a rate $\lambda_0$.
R and S symbolize two enantiomeric forms of amino acids
 in {\small L} and {\small D} forms.
We call them R and S chiral monomers respectively hereafter. 
Since the production of amino acids requires energy, 
the rate $k_0$ is very small. 
Only in some special environment and places, 
for instance under thunderstorms or close to seafloor hydrothermal vent,
the energy is supplied to increase $k_0$.
But if there is only the enhancement of random production by $k_0$, 
then the racemic mixture of R and S monomers results, as is shown
in the following.
Thus, for chirality selection, we have to consider further mechanism.

Chiral monomers are assumed to polymerize if they are of 
the same enatiomeric type as
\begin{align}
R + R_{i} ~ \udarrow{k_i}{\lambda_i}  R_{i+1} ,  \qquad 
S + S_{i} ~ \udarrow{k_i}{\lambda_i}  S_{i+1} ,   
\label{eq5}
\end{align}
with $i = 1, 2, \cdots$.
Here $k_i$ and $\lambda_i$ represent respectively the polymerization and
decomposition rate of the chiral polymers  R$_{i+1}$ or S$_{i+1}$. 
Generally, polymers at higher levels or macromolecules have some chemical 
functions.
When the $N$-mers, R$_N$ and S$_N$, act
as enzymes to reproduce the monomer of the same enatiomeric type,
the catalytic process is described as
\begin{align}
A+R_N \udarrow{k_e}{\lambda_e} R+R_N ,
\qquad
A+S_N \udarrow{k_e}{\lambda_e} S+S_N .
\label{eq6}
\end{align}
Here $k_e$ represents the rate of catalytic production
and  $\lambda_e$ that of back reaction.
A good enzyme should have a large production rate $k_e$ compared to $k_0$,
that of the non-autocatalytic production process.
However, the ratio of rate coefficients of a forward and a backward reactions
is specified by the energy difference of initial and final states, 
A and R (or S).
Then the back reaction with $\lambda_e$ should also be enhanced so as to satisfy
the following relation
\begin{align}
\frac{\lambda_e}{k_e} = \frac{\lambda_0}{k_0} .
\label{eq7}
\end{align}
For simplicity, further polymerization starting upward 
from the enzyme $N$-mers will not be 
considered in the following: namely, the reaction processes (\ref{eq5}) 
only up to $i \le N-1$ are considered here.

While the whole system is described by the reaction processes 
(\ref{eq4}-\ref{eq6})
 for a closed system, there are other processes for an open system;
  that is the exchange of substrate molecules between the system 
  and the environment E as
 is illistrated as 
\begin{align}
E \udarrow{F}{\mu_0} A,
\label{eq8}
\end{align}
and the leakage of polymers out of this system into the environment as 
\begin{align}
R_i \uarrow{\mu_i} E.
\label{eq9}
\end{align}
The leakage process does not necessarily mean the outflow
of the materials to the ambience
but can be consumptions of polymers for other utilities.

In the present paper, we consider the simplest case such 
that the dimer is sufficient to 
present catalytic effect, and assume $N=2$.
Then, the concentrations of achiral substrate $a$
and of chiral products up to the dimers
 $r,~r_2,~s,~s_2$ evolve respectively as
\begin{align}
\dot r =& k_0 a - \lambda_0 r +k_e ar_2 - \lambda_e r_2 r 
-2 k_1 r^2 + 2 \lambda_1 r_2 -\mu_1 r,
\nonumber \\
\dot r_2=& k_1 r^2 -\lambda_1 r_2  - \mu_2 r_2,
\nonumber \\
\dot s =& k_0 a - \lambda_0 s +k_e as_2 - \lambda_e s_2 s 
-2 k_1 s^2 + 2 \lambda_1 s_2- \mu_1 s,
\nonumber \\
\dot s_2=& k_1 s^2 -\lambda_1 s_2 - \mu_2 s_2 
\nonumber \\
\dot a =& -2k_0 a +\lambda_0(r+s) -k_e a(r_2+s_2) + \lambda_e (r_2 r + s_2 s)
\nonumber \\
& +F - \mu_0 a.
\label{eq10}
\end{align}
These rate equations yield that the total concentration of molecules
$c= a+r+2r_2+s+2s_2$ evolves according to 
\begin{align}
\dot c= F-\mu_0 a - \mu_1(r+s)-2 \mu_2(r_2+s_2).
\label{eq11}
\end{align}

Let us first consider a closed sytem with $F=\mu_i=0,\,(0\le i \le 2)$. 
Then $c$ remains constant $c=c_0$. 
Even with this conservation law, rate equations are still too complicated 
to study the chirality selection analytically, since there are
many variables coupled nonlinearly.
Therefore, we utilize a steady-state approximation
\cite{saito+05a}
such that the dimerization proceeds promptly and one may set 
$\dot r_2=\dot s_2=0$. 
Then the steady-state 
concentrations of dimers are determined in terms of monomers as
$r_2=({k_1}/{\lambda_1}) r^2$ and $s_2=({k_1}/{\lambda_1}) s^2$, and
the rate equations for the monomers reduce to
\begin{align}
\dot r = &(k_0+\frac{k_ek_1}{\lambda_1} r^2)(a-\frac{\lambda_0}{k_0} r)
\nonumber \\
\dot s = &(k_0+\frac{k_ek_1}{\lambda_1} s^2)(a-\frac{\lambda_0}{k_0} s)
\label{eq12}
\end{align}
with $a=c_0-(r+s)-2(r_2+s_2)$.
Here the relation eq.(\ref{eq7}) is used.
Dimer enzymes lead to nonlinear autocatalytic effect,
but the effect is 
compensated
 by the same amount of enhanced back reaction.
Since $k_0+(k_ek_1/\lambda_1) r^2$ and $k_0+(k_ek_1/\lambda_1) s^2$ 
are always positive, the 
sole
 possible fixed point is racemic:
$\bar r=\bar s=(k_0/\lambda_0) \bar a$.
Only in the absence of the back reaction, $\lambda_0=\lambda_e=0$, the
system has a fixed line $a=0$.
There is no chirality selection if the system has a fixed line,
as was pointed out in the previous studies of a closed system.
\cite{saito+04a,saito+04b}

From the above analysis, it is concluded that the polymerization system 
(our model at least) has to
be open for the chiral symmetry breaking to occur. 
To analyse the open system theoretically, we again use the 
steady-state approximation
 by replacing the dimer concentrations by their stationary values,
$r_2=k_1 r^2/(\lambda_1+\mu_2),~s_2=k_1 s^2/(\lambda_1+\mu_2)$. 
The approximation does not alter the structure of the fixed points
for the system.
The reduced rate equations for monomers are obtained as
\begin{align}
\dot r = &(k_0+\frac{k_ek_1}{\lambda_1+\mu_2} r^2)a 
\nonumber \\
&
- (\lambda_0+\frac{\lambda_ek_1}{\lambda_1+\mu_2} r^2 +
 \mu_1 + \frac{2 \mu_2 k_1}{\lambda_1+\mu_2} r) r,
\nonumber \\
\dot s = &(k_0+\frac{k_ek_1}{\lambda_1+\mu_2} s^2)a 
\nonumber \\
&
- (\lambda_0+\frac{\lambda_ek_1}{\lambda_1+\mu_2} s^2+ 
\mu_1 + \frac{2 \mu_2 k_1}{\lambda_1+\mu_2} s)s.
\label{eq13}
\end{align}
The enantiomeric excess order parameter
of the monomer is defined by
\begin{align}
\phi_1 = \frac{r-s}{r+s}.
\label{eq14}
\end{align}
From eq.(\ref{eq13}) it is shown to evolve according to 
\begin{align}
\dot \phi_1 = A \phi_1 - B \phi_1^3
\label{eq15}
\end{align}
with 
\begin{align}
&A= B- \frac{2k_0a}{r+s}, 
\nonumber \\
&B=
\frac{k_1(r+s) }
{2(\lambda_1+ \mu_2)}[ak_e-\lambda_e(r+s) - 2\mu_2].
\label{eq16}
\end{align}
One notices that $\lambda_0,~\mu_0$ and $\mu_1$ do not appear 
explicitly in eq.(\ref{eq16}), but their effects are implicitly included in
the values of concentrations, $a$ and $r+s$.

As long as the concentrations $a$ and $r+s$ are positive, 
the coefficient $B$ is always larger than $A$.
If $A$ is positive 
asymptotically or at a fixed point, then $\phi_1$
approaches to a nonzero value as $\phi_1^2=A/B$ and 
the chiral state will be
selected. 
For example, if $A>0$ at racemic fixed points ($\bar r=\bar s)$,
they are unstable and the system may evolve to chiral fixed points.
On the contrary, if $A<0$ at racemic fixed points, they are stable.
For $\mu_0=\mu_2=0$, the racemic fixed point is easily obtained
from eq.(\ref{eq10}) and eq.(\ref{eq11}) as 
\begin{align}
\bar r=\bar s= \frac{F}{2 \mu_1}, \quad 
\bar a= \frac{\mu_1 + \lambda_0 + (\lambda_e k_1/\lambda_1) \bar r^2}
{k_0+(k_e k_1/\lambda_1) \bar r^2} \bar r .
\label{eq17}
\end{align}
By using the relation eq.(\ref{eq7}), the coefficient $A$ in eq.(\ref{eq16})
is written as
\begin{align}
&A=- k_e 
\nonumber \\
& \times\frac{\lambda_e (k_1 \bar r^2/\lambda_1)^2 -  (\mu_1- 2 \lambda_0)
 (k_1 \bar r^2/\lambda_1) +\lambda_0(\lambda_0+\mu_1)/\lambda_e}
 {k_e (k_1 \bar r^2/\lambda_1) +k_0}.
 \label{eq18}
\end{align}
Therefore, $A$ is positive and the racemic fixed point is unstable,
if the influx $F$ is in the range 
between the lower bound $F_1$ and the upper one $F_2$,
 given as 
\begin{align}
F_{1,2} = \mu_1 \sqrt{
\frac{2 \lambda_1}{\lambda_e k_1}\Big(
\mu_1-2 \lambda_0 \mp \mu_1\sqrt{1-8 \lambda_0/\mu_1}
\Big)}
\label{eq19}
\end{align}
with $\mu_1>8 \lambda_0$. 
This region where the racemic fixed point is unstable is
denoted by "chiral" in the $F - \lambda_0/k_0$ phase space 
in Fig.1(a). 
Parameters are fixed at 
$k_0=0.1,~k_1=1,~k_e=10$ as well as at $\lambda_1=\mu_1=1$, and
the decomposition rates back to achiral substrate $\lambda_0,\lambda_e\,$ are 
varied by keeping the relation (\ref{eq7}) satisfied.
At a very large $F$, the racemic fixed point recovers stability, since
$\bar r= \bar s$ increase and the
chirality suppression factor $\lambda_e(\bar r+ \bar s)$ in $B$ 
due to a finite back reaction rates $\lambda_{0,e}$
exceeds the chirality enhancement factor 
$\bar a k_e \approx \lambda_e \bar r$.
Therefore, by decreasing the back reaction,
the racemic fixed point is expected to remain unstable at large $F$.
Actually, at $\lambda_0=\lambda_e=0$, the upper bound $F_2$ diverges and the
instability of racemic fixed point takes place for all $F$'s larger 
than the lower bound 
\begin{align}
F_1 = 2 \mu_1 \sqrt{\frac{ \lambda_1 k_0}{k_1 k_e} }.
\label{eq20}
\end{align}

\begin{figure*}[h]
\begin{center} 
\includegraphics[width=0.43\linewidth]{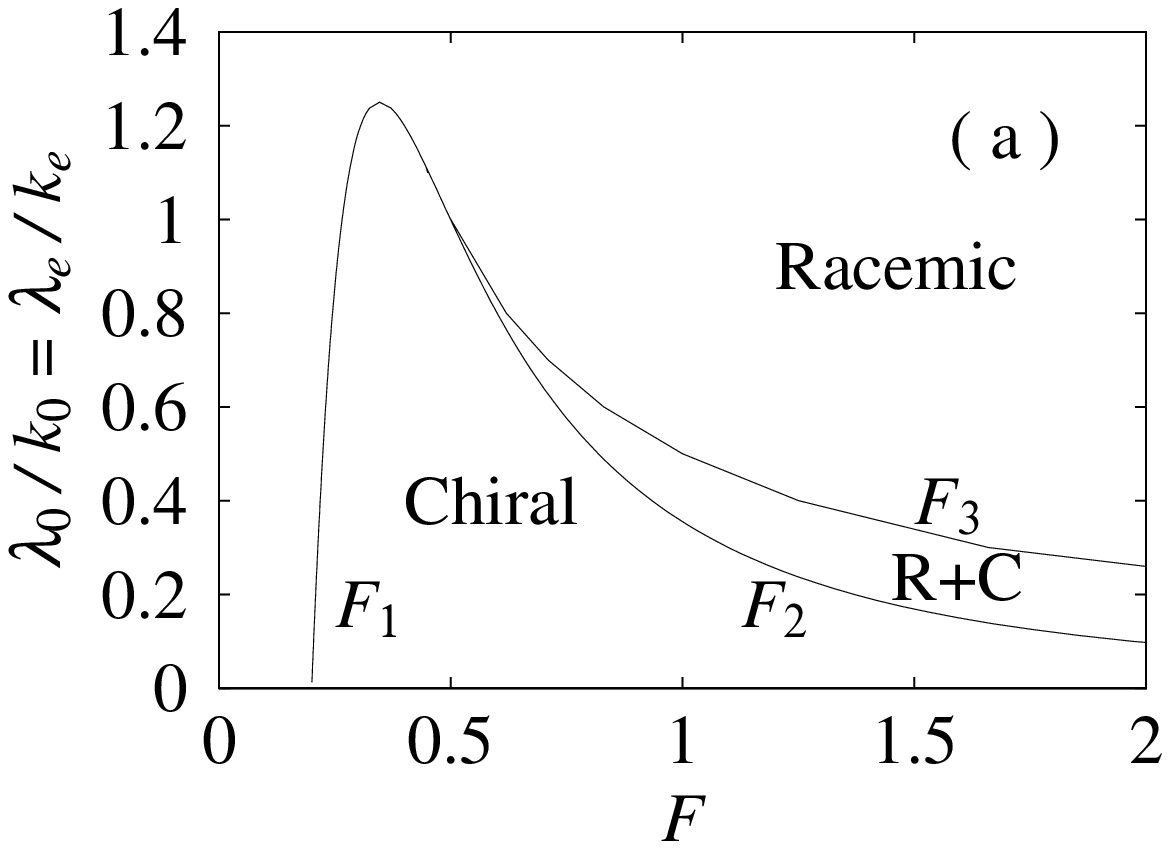}
\includegraphics[width=0.43\linewidth]{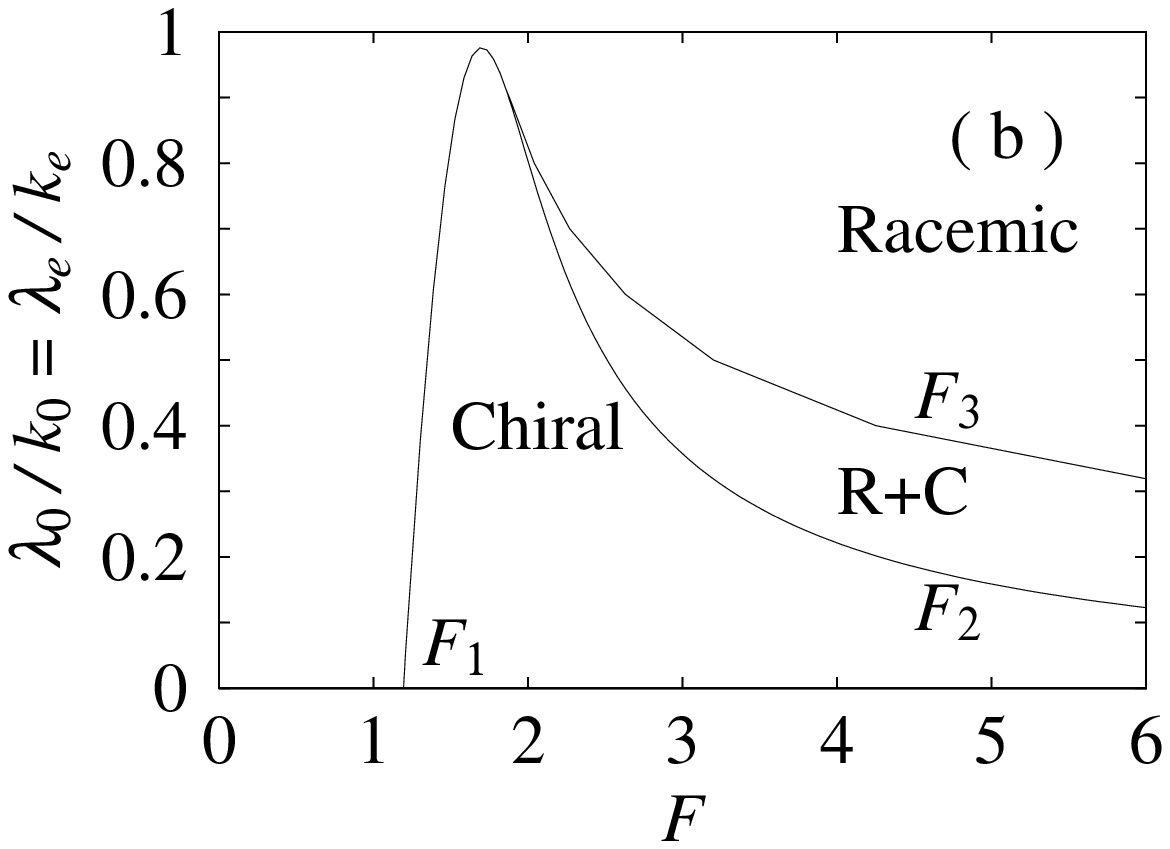}
\end{center} 
\caption{Phase boundary between the racemic and the chiral fixed points
in the $F - \lambda_0/k_0$ phase space.
Production parameters are fixed at $k_0=0.1,~k_1=1,~k_e=10$.
Other parameters fixed are (a) $\mu_0=\mu_2=0$, $\lambda_1=\mu_1=1$, and (b) 
$\mu_{0,1,2}=\lambda_1=1$.
}
\label{fig1}
\end{figure*}

When the back reaction is too strong as $\lambda_0> \mu_1/8$,
or when the substrate influx is too strong as $F \gg F_2$,
there remain plenty of achiral substrates and the random production
by $k_0$ dominates over the catalytic production with $k_e$.
When the substrate influx is too weak as $F <F_1$, 
there are too few product
chiral molecules to sustain catalytic chiral enhancement.
In those regions, racemic fixed point is the only possibility.
With a small decomposition $\lambda_0 < \mu_1/8$ and with a moderate
substrate influx, 
states with broken chirality are possible.

\begin{figure*}[h]
\begin{center} 
\includegraphics[width=0.32\linewidth]{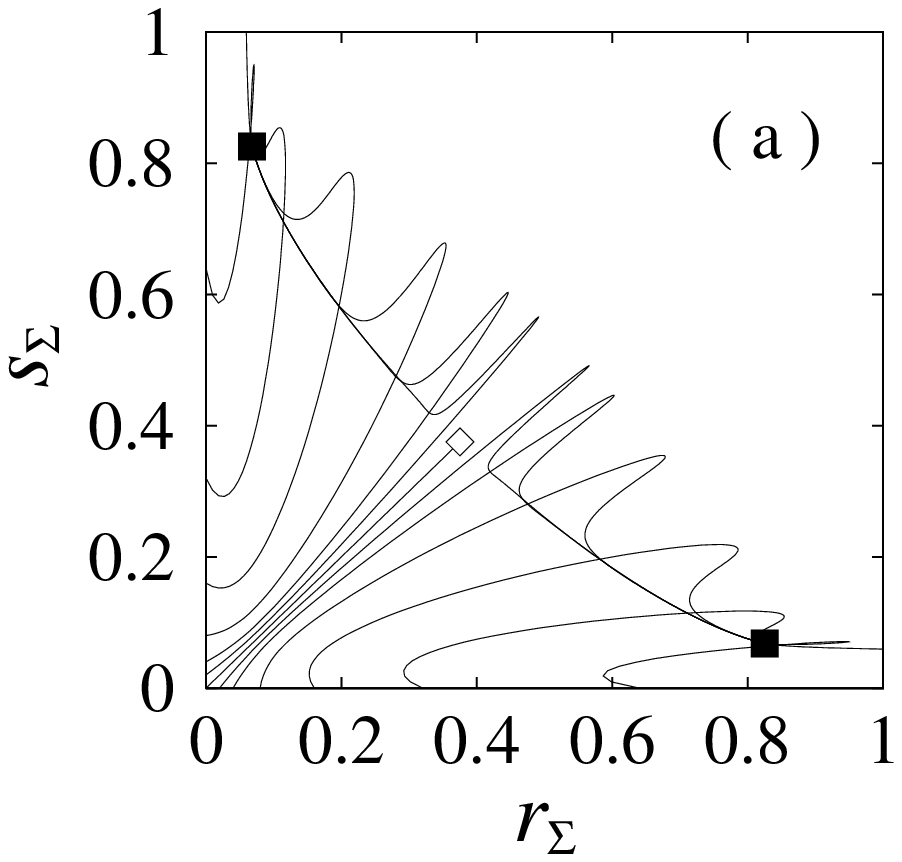}
\includegraphics[width=0.32\linewidth]{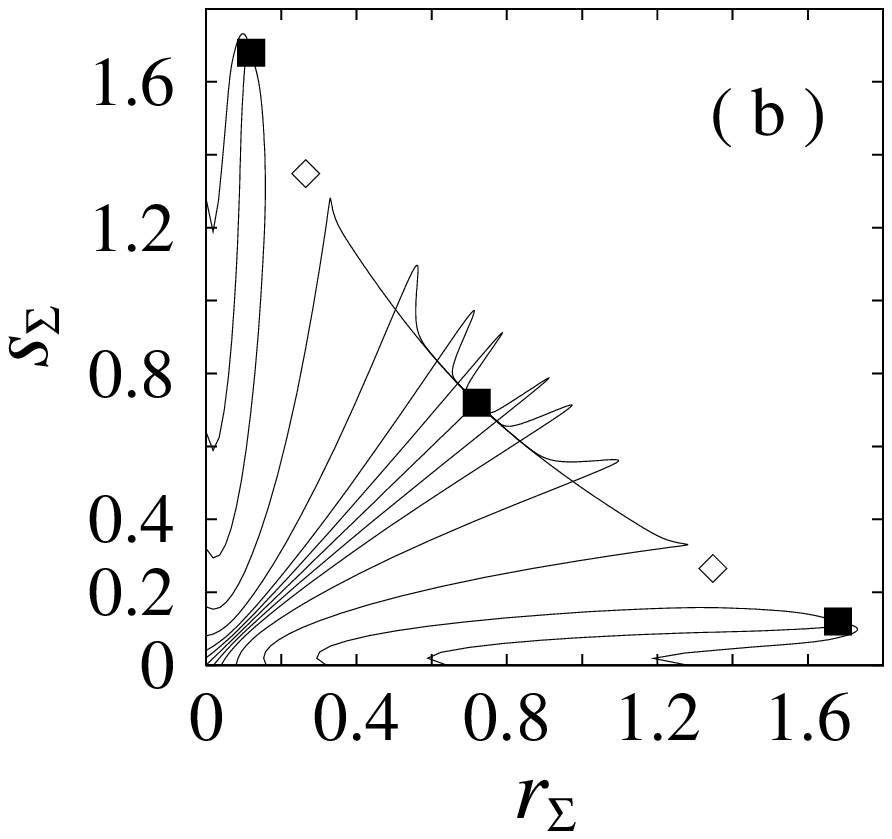}
\includegraphics[width=0.32\linewidth]{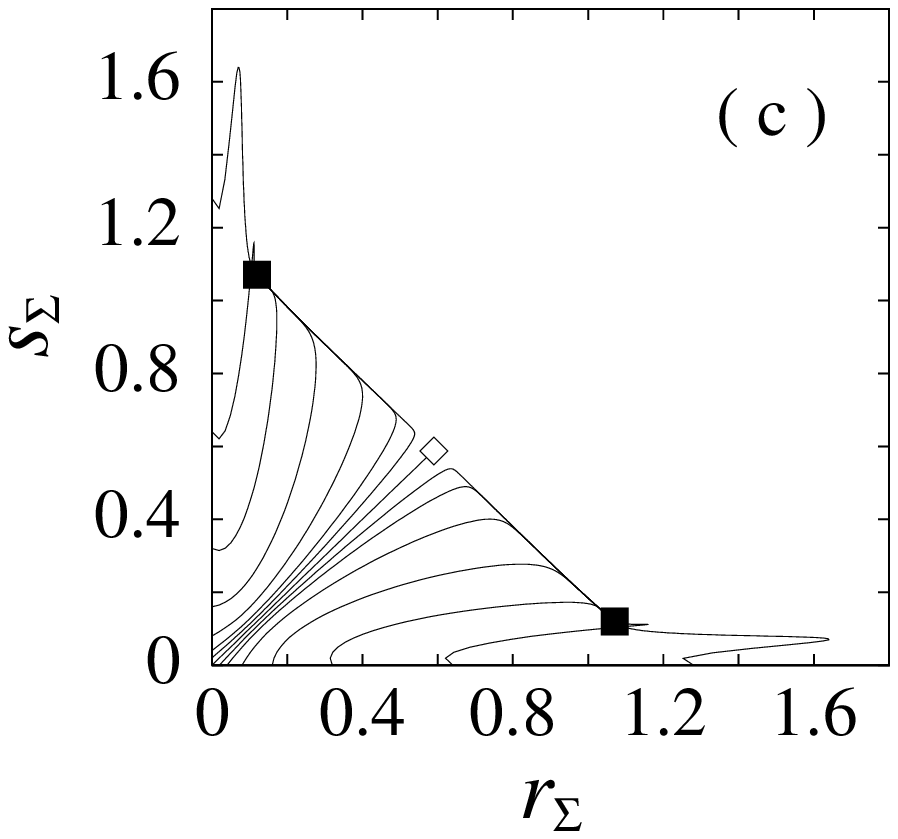}
\end{center} 
\caption{Flow diagram in the total concentrations of $r$ and $s$ phase space.
Black boxes represent stable fixed points, and white diamonds
unstable ones. 
Production parameters are fixed at $k_0=0.1,~k_1=1,~k_e=10$.
Figures (a) and (b) correspond to Fig.1(a) with decomposition rates 
$\lambda_1=1, \lambda_0/k_0 =\lambda_e/k_e =0.6 $ and flow $\mu_0=\mu_2=0$.
The in-coming flux of the substrate is (a)$F=0.5$ and (b) $F=0.8$.
(c) corresponds to Fig.1(b) with decomposition rates
$\lambda_1=1, \lambda_0/k_0 =\lambda_e/k_e =0.5 $ and flow 
$\mu_0=\mu_1=\mu_2=1$.
The in-coming flux of the substrate is $F=2$.
}
\label{fig2}
\end{figure*}

We note that the conclusions of the instability on the racemic fixed point
remain valid for the full system, irrespective of the approximations. 
Equipped with these results, we carry out numerical integration 
of the full rate equations (\ref{eq10}) systematically. 
All concentrations except $r$ (or $s$) are set zero initially, and the
initial value of $r$ (or $s$) is varied to obtain the flow trajectories.
The flow diagram in the phase space of 
the total concentrations of two enantiomers, $r_{_{\Sigma}}=r+2r_2$
and $s_{_{\Sigma}}=s+2s_2$, shows that the chiral symmetry breaking 
actually takes place in the region denoted by "chiral" in Fig.1(a). 
For example, flow trajectories at $F=0.5$ with 
$\lambda_0/k_0=\lambda_e/k_e=0.6$ in Fig.2(a) clearly shows 
the symmetry breaking: Fixed points with broken chiral symmetry attract
all the trajectories.

\begin{figure*}[h]
\begin{center} 
\includegraphics[width=0.45\linewidth]{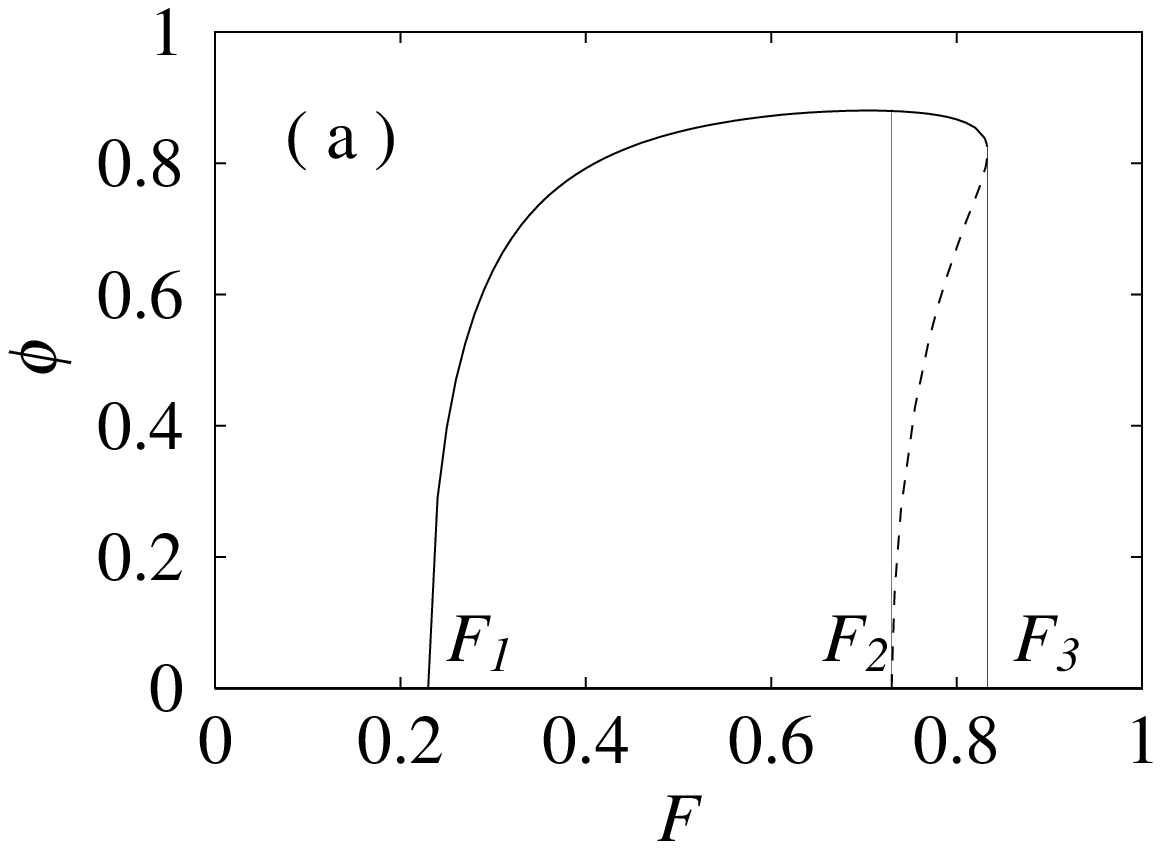}
\includegraphics[width=0.45\linewidth]{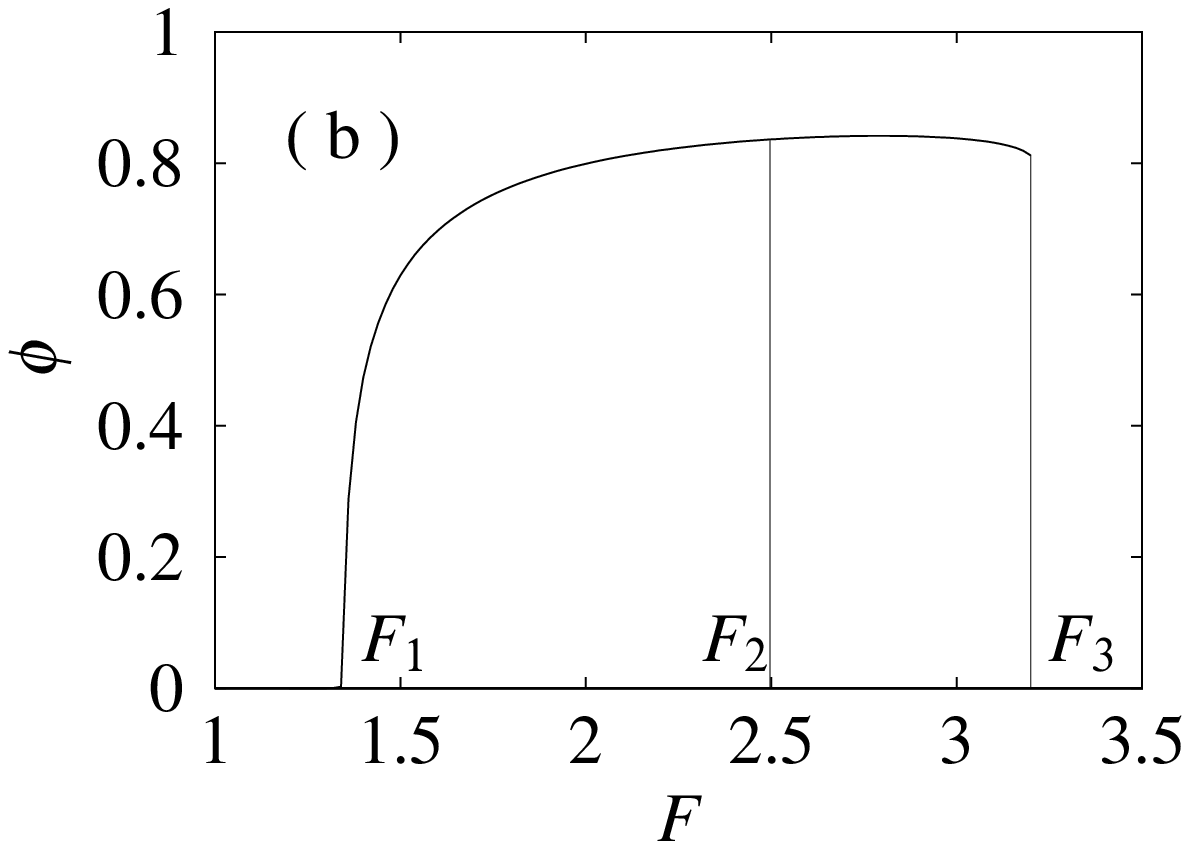}\\
\includegraphics[width=0.45\linewidth]{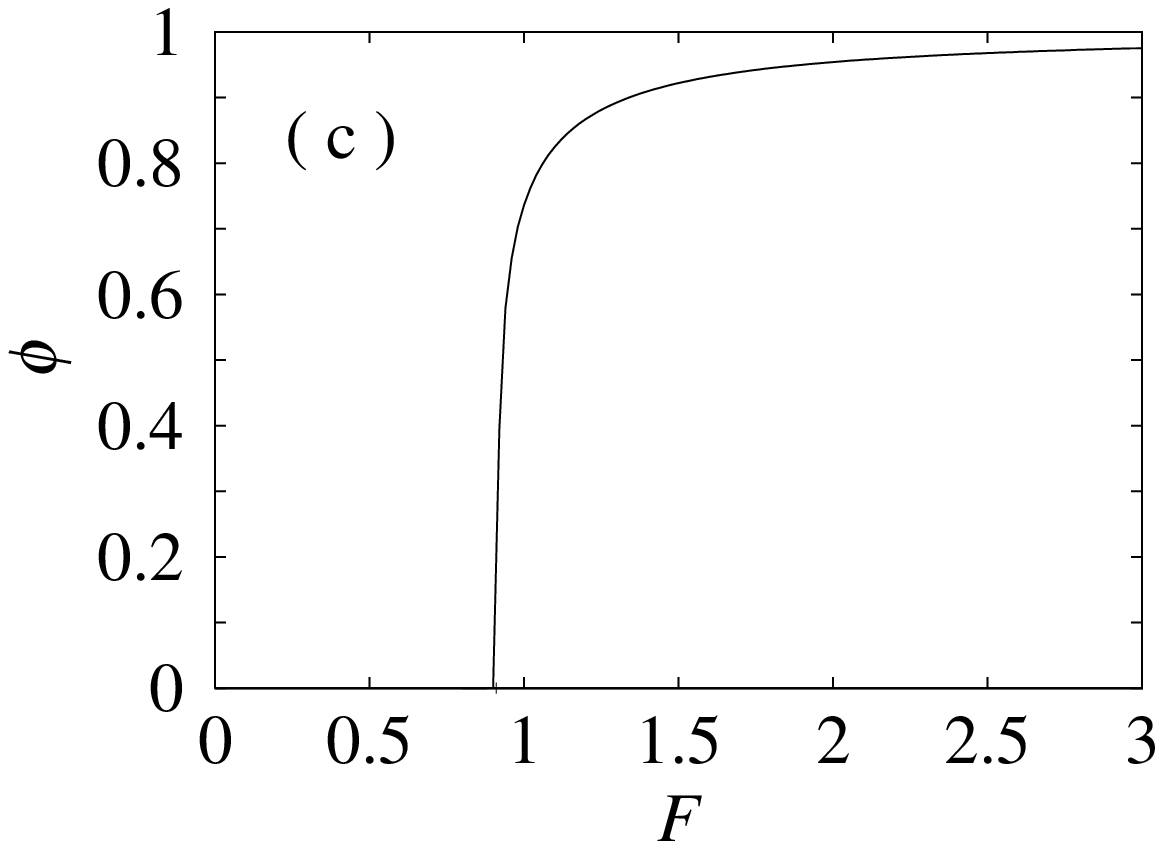}
\includegraphics[width=0.45\linewidth]{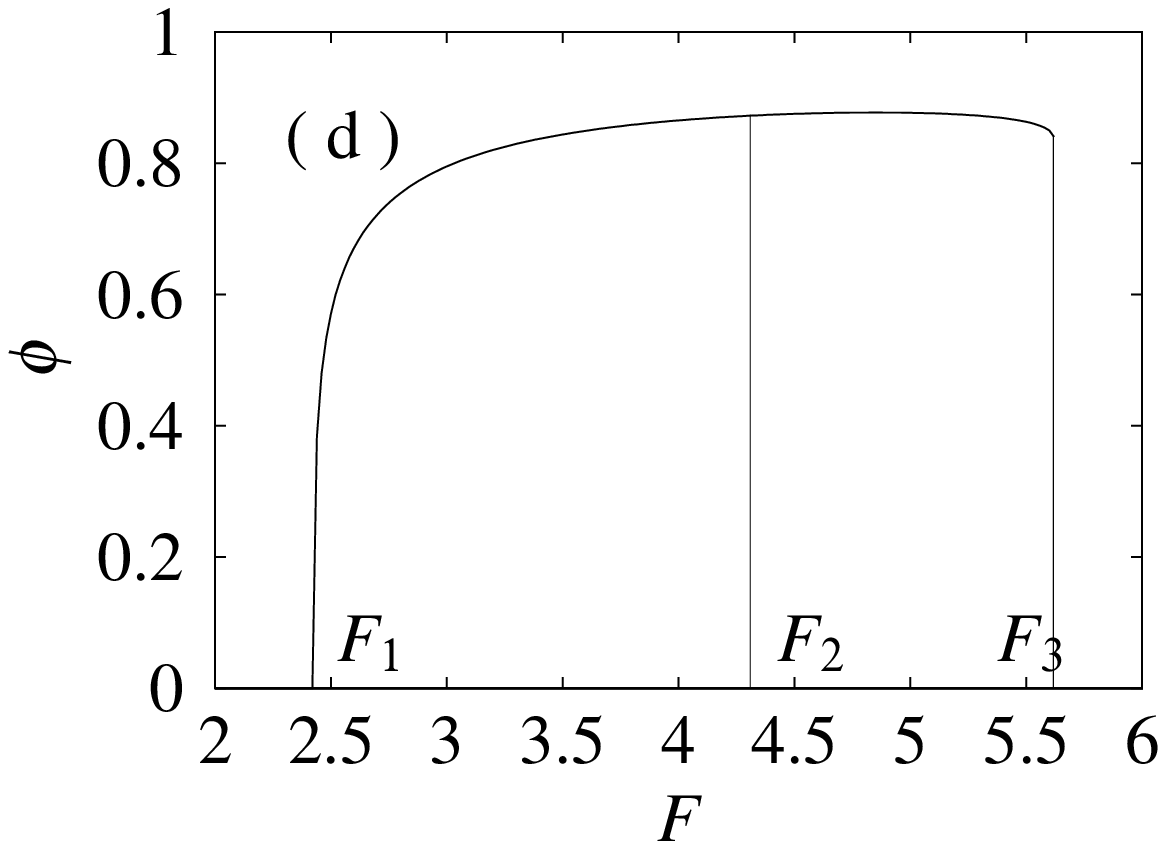}
\end{center} 
\caption{Fixed point value of the total enantiomeric excess $\phi$ 
versus influx $F$.
Production parameters are fixed at $k_0=0.1,~k_1=1,~k_e=10$.
Figure (a)  
corresponds to Fig.1(a) with decomposition rates 
$\lambda_1=1, \lambda_0/k_0 =\lambda_e/k_e =0.6 $ and out-flow 
$\mu_1=1,~\mu_0=\mu_2=0$.
(b)  
corresponds to Fig.1(b) with decomposition rates 
$\lambda_1=1, \lambda_0/k_0 =\lambda_e/k_e =0.5 $ and out-flow 
$\mu_0=\mu_1=\mu_2=1$.
(c) 
corresponds to Fig.4(a) without decompositions,
$\lambda_i=0$ and out-flow 
$\mu_0=\mu_1=\mu_2=1$.
(d) 
corresponds to FIg.4(b) with decompositions
$\lambda_i=k_i$ and out-flow 
$\mu_0=\mu_2=1, ~\mu_1=2$.
}
\label{fig3}
\end{figure*}

The degree of chiral symmetry breaking is measured by the total enantiomeric
excess 
\begin{align}
\phi=\frac{r_{_{\Sigma}}-s_{_{\Sigma}}}{r_{_{\Sigma}}+s_{_{\Sigma}}}.
\label{eq21}
\end{align}
Within the chiral region, the fixed point value of $\phi$ varies as the 
incoming flux $F$, as shown in Fig.3(a).
Due to the symmetry, there is a trivial symmetric branch with negative $\phi$ 
values, which is not shown.
Near above the lower bound $F_1$, the ee increases continuously.
Here, chiral fixed points lie close to the racemic one, 
and the parameter $A$ in eq.(\ref{eq16}) is approximately written as
$A \propto (F-F_1)(F_2-F)$. Since the $F$ and $\bar r$ are small, contributions
of dimers $r_2$ in $r_{_{\sum}}$ is negligibly small. Thus
$\phi \approx \phi_1 \propto \sqrt{F-F_1}$.

As $F$ increases, $\phi$ increases as well, and the chiral
fixed point extends for $F$ larger than $F_2$, where the racemic fixed point
recovers its stability.
In fact, for F between $F_2$ and $F_3$ there are two types of stable
fixed points, racemic and chiral, in the flow diagram, as shown in 
Fig.2(b) at $F=0.8$.
If the initial concentrations $r$ and $s$
 for the numerical integration of
eq.(\ref{eq10}) 
are close to the racemic configuration, the final state remains racemic, since
it is stable, at least locally, as proven by the
stability analysis.
However, if the integration start from a well developped chiral configuration, 
the chiral fixed point is selected.
The concentration phase space of $r$'s and $s$'s is thus devided into multiple
basins of attractions to racemic and chiral fixed points.
When $F$ is larger than $F_3$, the basins of attraction
to the chiral fixed points seem to disppear.

From Fig.2(b) it is evident that  
for $F$ between $F_2$ and $F_3$ there are unstable chiral fixed points
(marked by white diamonds)
between the racemic and stable chiral fixed points (marked by black boxes).
Therefore, in the $F~-~\phi$ diagram in Fig.3(a), there is 
another branch 
corresponding to these unstable chiral fixed points, denoted by a
dashed curve:
it extends 
from $F_3$ back to the small $F$ direction below the stable
chiral branch. 
The unstable branch seems to be connected to the $F$ axis at a point $F_2$.

The maximum value of ee in Fig. \ref{fig3} is rather small.
This is due to the present choice  of a parameter
set, especially $k_0$, which is chosen rather large on purpose 
so that the fixed points are clearly discernible  in
figures. With a parameter set in Fig.2(a), for example, the ee takes a value
$\phi=0.871$, but with a smaller value of $k_0$ as $k_0=0.01$,
the ee increases up to $\phi=$0.992 and the stable chiral fixed points
lie almost on two axes.

Every aspect of the phase boundary and the flow trajectories
remains the same, even
if all the out-flow rates $\mu_i$ take  finite values, 
for example $\mu_0=\mu_1=\mu_2=1$:
Phase diagram in $F$ versus $\lambda_0/k_0$ in Fig.1(b)
looks similar to that in Fig.1(a).
When $F$ takes the value between $F_1$ and $F_2$, 
the racemic fixed point is unstable and
the chiral symmetry breaks as shown in Fig.2(c).
Between $F_2$ and $F_3$, there is a hysteresis such that the 
stationary state depends on the initial concentrations of
$r$'s and $s$'s.
From the lower bound $F_1$ the ee increases continuously until it
drops suddenly to zero at $F_3$, as shown in Fig.3(b), where
the expected unstable chiral fixed points are omitted.

We now consider cases without any direct back reaction, $\lambda_i=0$
for $i=1$ to 3. Still, 
part of chiral products is utilized for 
other purposes and disappears from the catalyst production route, such that
$\mu_i$'s are finite. 
By fixing $\mu_0=\mu_2=1$ and at various $\mu_1$, the chiral symmetry 
breaking is achieved at an incoming flux $F$ larger than the lower bound,
as shown in Fig.4(a). 
One notices that there is a minimum in the lower bound of the
incoming flux $F$.
With $\mu_0=\mu_1=\mu_2=1$, the ee increases rapidly 
to the saturation  value $\phi$ close to unity, as shown in Fig.3(c).
With small $k_0$ and large $k_e$, the saturation to $\phi=1$
happens almost instantaneously.

\begin{figure*}[h]
\begin{center} 
\includegraphics[width=0.45\linewidth]{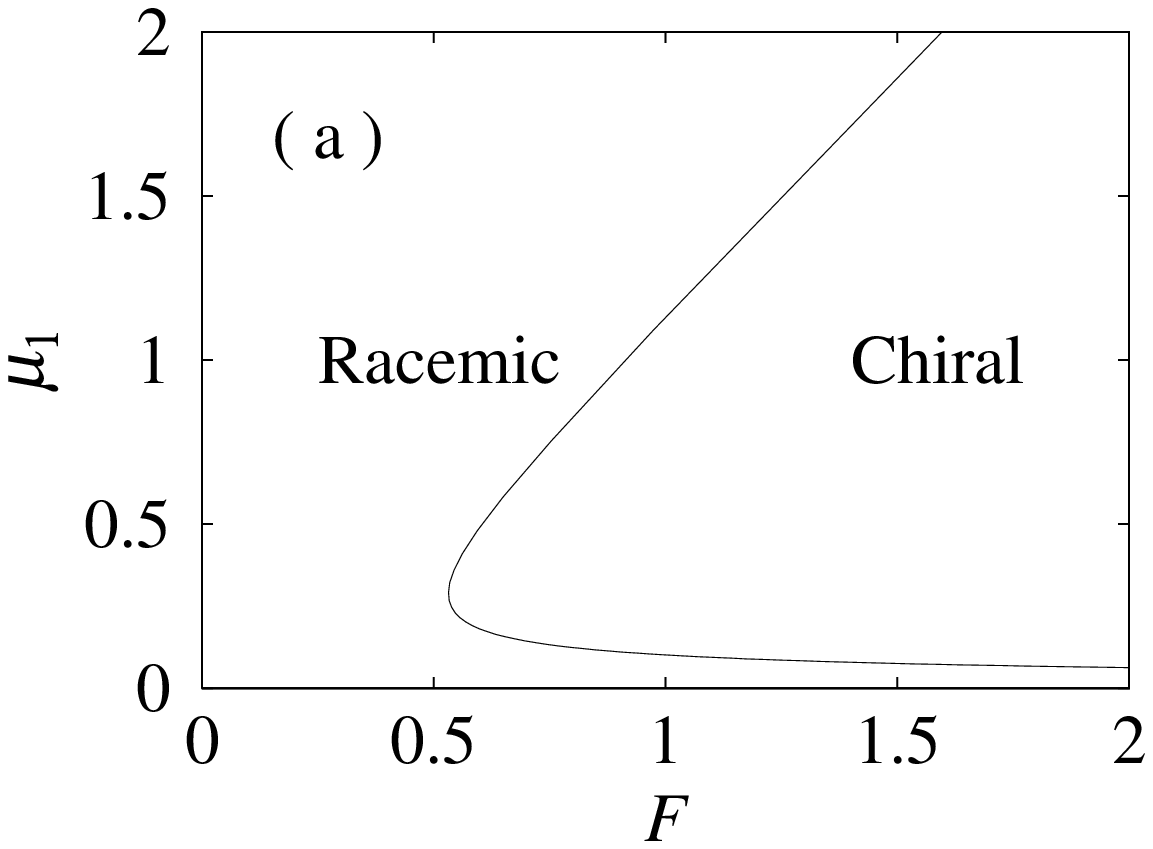}
\includegraphics[width=0.45\linewidth]{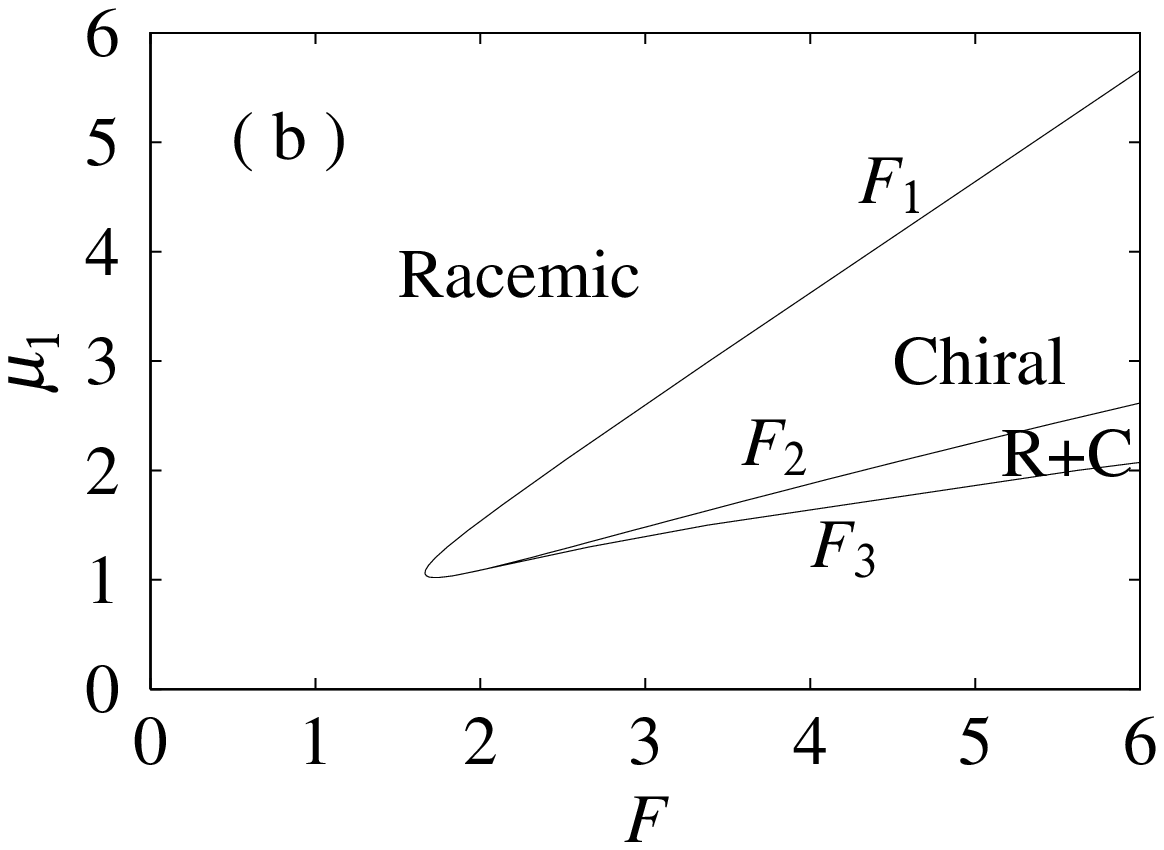}
\end{center} 
\caption{Phase boundary between the racemic and the chiral fixed points
in the $F - \mu_1$ phase space.
Production parameters are fixed at $k_0=0.1,~k_1=1,~k_e=10$, 
and $\mu_0=\mu_2=1$.
Rates of back reactions are (a) $\lambda_i=0$,  and (b) 
$\lambda_i/k_i=1$ with $i=0,~1,~e$.
}
\label{fig4}
\end{figure*}

With finite rates of back reactions, as $\lambda_i=k_i$ for example,
the out-going $\mu_1$ has to be sufficiently large to realize the chirality
selection, as shown in Fig.4(b).
The symmetry breaking is possible for $F$ between $F_1$ and $F_3$
with hysteresis above $F_2$, similar to the previous cases
 in Fig.1(a) and (b).
The ee increases continuously from zero at the lower bound $F_1$,
and drops abruptly to vanish at the upper bound $F_3$ with hysterisis
above $F_2$, as shown in Fig.3(d).

\section{Summary and Discussions}

We extend our model of chiral molecule production from achiral substrate
with nonlinear autocatalysis to an open system under flow.
Even if the back reaction is not explicitly included, 
 the system after a transient period is shown to be effectively described
by the rate equations we derived for a closed system with back reaction, 
and the chirality selection is expected to take place. 
As an ideal system to realize homochirality in laboratory, 
we proposed the Soai reaction under flow.

Organic molecules in life, amino acids for instance, may not undergo
nonlinear autocatalytic reaction, but they polymerize and acquire
some functions. One of them is to catalyze various chemical reactions,
presumably the monomer production process as well.
We construct a model of chiral monomer production from achiral
substrate with further polymerization to produce enzymes which enhance
monomer production of the same enatiomeric type.
By assuming for simplicity that dimers already act as enzymes, 
the rate equations of monomer concentrations
reduce to those with quadratic autocatalysis in a steady-state approximation.
In a closed system, however, 
the enzyme enhances the decomposition process as well, and the chiral symmetry
is not broken.
Only in an open system where substrate is supplied and parts of
produced monomers and polymers are consumed for other purposes, the
chiral symmetry can be broken if parameters take proper values.
Influx of substrate should be large to yield sufficient chiral products
to sustain catalytic process, but if the back reaction takes place
the influx should not be too large. Otherwise, an enhanced back reaction
compensates the chirality enhancement. 

We have considered only a single chiral species for simplicity reason, 
whereas in reality there are 20 types of amino acids.
If only the same enantiomeric type can form stable polymers,
we believe that the present mechanism of chirality selection is still valid
for multiple chiral species.

By extending the ladder for polymerization processes 
until the enzyme is produced, 
the window in parameter space for the chiral symmetry breaking becomes
narrower. 
Various disturbances due to spatial and temporal variations may tend to narrow
this window further. 
However narrow the parameter window may be, 
it is essential that the window is open.
After all, it seems to be a majority concensus
 that several hundred million years were needed for the birth of life on the 
 earth.

\acknowledgement
Authors acknowledge support from the Gakuji-Shinkou-Shikin 
by Keio University.

\appendix
\section{}

A general set of the (reduced) rate equations for the concentrations $r,s,a$ is 
\begin{align}
\dot r&=f(r)a-\lambda(r)r\nn
\dot s&=f(s)a-\lambda(s)s  \label{eqA1}\\
\dot c&=g(a)-h(r)-h(s)\nn
c&=a+j(r)+j(s)\nonumber
\end{align}
where $f(r)$ and $\lambda(r)$ are the effective coefficients of the production reaction and the back reaction, respectively. $c$ is the total concentration and
 functional forms of $g(a),h(r),j(r)$ depend on models but presumably are nonnegative functions. An enantiomeric excess order parameter defined as
\begin{align}
\phi=\frac{r-s}{r+s}
\end{align}
is shown from eq. (\ref{eqA1}) to evolve according to
\begin{align}
\dot \phi&=A\phi-B\phi^3
\end{align}
with
\begin{align}
A&=B-\frac{2f(0)a}{r+s}\\
B&=\frac12(r+s)
\nonumber \\
& \times \Bigl\{\frac{s(f(r)-f(0))-r(f(s)-f(0))}{
rs (r-s)}a-\frac{\lambda(r)-\lambda(s)}{r-s}\Bigr\}
\nonumber\end{align}
The form of the parameter $B$ is so chosen that it remains finite for
the racemic state $r=s$ as well as at the trivial point
$r=s=0$.
For specific form $f(r)=k_0+k_1r+k_2 r^2 $ and 
$\lambda(r)=\lambda_0+\lambda_1r+\lambda_2r^2$, $B$ becomes simply
\begin{align}
B=\frac12(r+s)\bigl\{k_2a-\lambda_1-\lambda_2(r+s)\bigr\}.
\end{align}
We should not forget that $A$ and $B$ are functions of the variable $r,s,a$
 and varies as functions of time accodingly. 
 
For simplicity, let us assume that $\bar r\bar s\bar a\ne 0$ at a fixed point
 $r=\bar r,~s=\bar s, ~a=\bar a$, which satisfy 
 \begin{align}
 f(\bar r)\bar a=\lambda(\bar r)\bar r,\quad f(\bar s)\bar a=\lambda(\bar s)\bar s. \end{align}
 Using these equations, one obtains
 \begin{align}
 \bar B=\frac12 (\bar r+\bar s)\frac{f(0)\bar a}{\bar r\bar s}
 \end{align}
and $ \bar A=\bar{\phi}^2\bar B$. The last relation only shows the consistency.
To analyse the stability of the fixed point, one should first to obtain the
fixed point values of $a,r,s$ as functions of relevant parameters and
finds the conditions for the chiral phase $A> 0$ or for the racemic phase
$ A\le 0$ as is discussed in section 3. 

The other way to analyse the stability is the linear stability
analysis. As a concrete example, we consider the stability matrix for the 
full 
system eq.(\ref{eq10}) at the racemic fixed point eq.(\ref{eq17}) supplimented with 
$\bar{r}_2=k_1 \bar{r}^2/(\lambda_1+\mu_2),\,\bar{s}_2=k_1 \bar{s}^2/(\lambda_1+\mu_2)$. 
Since there are five independent variables as $(x_1,x_2,x_3,x_4,x_5)\equiv (r,r_2,s,s_2,a)\,$, the linear deviation around a fixed point $\bar{x}_i$ is assumed to be $x_i=\bar{x}_i+x_{i0}\exp(\Omega t)$. The equation to determine the eigenfrequency $\Omega$ is an eigenvalue equation $|\Gamma -\Omega I|=0$,
where $I$ is the $5\times 5$ dimensional unit matrix and the matrix elements of $\Gamma$ are given as $\Gamma_{ij}=\partial \dot{x}_i/\partial x_j$ evaluated at
the fixed point. In a matrix form, it is expressed as
\begin{align}
&\mbox{det}
\nonumber \\
&\begin{pmatrix}-S-\Omega&T&0&0&U\\ 2k_1\bar r&-\lambda_1-\Omega &0&0&0\\ 0&0& -S-\Omega&T&U\\ 0&0&2k_1\bar r& -\lambda_1-\Omega&0\\V&W&V&W&-2T-\Omega\end{pmatrix}
\nonumber \\
&=0
\end{align}
where the nonzero matrix elements are given as
\begin{align}
&S=\mu_1+\lambda_0+4k_1\bar r+\lambda_e \bar{r}_2\nn
&T=2\lambda_1+k_e\bar a-\lambda_e\bar r\nn
&U=k_0+k_e\bar{r}_2\\
&V=\lambda_0+\lambda_e\bar{r}_2\nn
&W=-k_e\bar a+\lambda_e\bar r.\nonumber
\end{align}
This equation is expressed as a product of a quadratic and a cubic equations as
\begin{align}
&\bigl[\Omega^2+(S+\lambda_1)\Omega+S\lambda_1-2k_1T\bar r\bigr]\nn
&\times \bigl[-(\Omega+S)(\Omega+\lambda_1)(\Omega+2U)
\nonumber \\
&+2k_1T\bar r(\Omega+2U)
+2UV(\Omega+\lambda_1)+4UWk_1\bar r\bigr]=0.
\end{align}
The instability condition for the racemic fixed point is the existence of
positive eigenfrequency. As $S+\lambda_1>0$ is always satisfied, an inequality $S\lambda_1-2k_1T\bar r<0$
is just the condition for the quadratic equation to have a positive root
and it can be shown that this condition agrees with that given in section 3.
As for the cubic equation, its roots are near $-S<0,\,-\lambda_1<0,\,-2T<0\,$
and thus are expected to be negative. Numerically we confirmed this expectation for reasonable parameter ranges.


\end{document}